\documentstyle[aps,epsf]{revtex}
\begin{document}
\preprint{UAHEP 977}
\draft
\title{Statistical mechanics of Kerr-Newman dilaton black holes \\
and the bootstrap condition}
\author{R. Casadio, B. Harms and Y. Leblanc}
\address{Department of Physics and Astronomy,
The University of Alabama\\
Box 870324, Tuscaloosa, AL 35487-0324}
\maketitle
\begin{abstract}
The Bekenstein-Hawking ``entropy'' of a Kerr-Newman dilaton black hole 
is computed in a perturbative expansion in the charge-to-mass ratio.
The most probable configuration for a gas of such black holes is 
analyzed in the microcanonical formalism and it is argued that it does 
not satisfy the equipartition principle but a bootstrap condition.
It is also suggested that the present results are further support for
an interpretation of black holes as excitations of extended objects.
\end{abstract}
In Refs. \cite{hl1,hl2} we analyzed gases of black holes and black
extended objects with various geometries.  
We have recently solved the field equations of the Kerr-Newman dilaton 
black hole in the small charge-to-mass approximation \cite{chlc}.
In this letter we show an attempt to use these solutions to calculate 
the quantum degeneracy of such a black hole and from it obtain the 
microcanonical density of states for a gas of such black holes.
We also improve previous analyses of the gas of Reissner-Nordstr\"om and
Kerr-Newman black holes.
\par
We assume that $\rho_{BH}$, the quantum degeneracy of a black hole, 
is given by the inverse probability ($T^{-1}$) for a particle of 
tunneling out the horizon.
In general one has $T\sim \exp(-S_E)$, where $S_E$ is the Euclidean 
action of the black hole.
However, it is known that the Euclidean action of a rotating black hole
does not exist and the calculation of the tunnelling probability
must invoke some suitable prescription for getting meaningful
results.
This has been given in Ref.~\cite{gh} for a Kerr-Newman black hole
and amounts to $\rho_{BH}\sim c\,\exp(A_{BH}/4)$, 
where $A_{BH}/4$ is the Bekenstein-Hawking ``entropy'' ($A_{BH}$ is the 
surface area of the horizon) and $c$ represents the quantum 
field theoretic effects (here we neglect possible non-local 
contributions).
Note that, in our interpretation \cite{hl1,hl2}, $A_{BH}/4$ is given
no statistical mechanical attribute.
\par
The Einstein-Hilbert action of dilatonic black holes is given by
($G=1$),
\begin{equation}
S_{BH}={1\over 16\,\pi}\int d^4x \,\sqrt{-g}\,
\left[R-{1\over 2}\,(\nabla\phi)^2-e^{-a\,\phi}\,F^2\right]+
\Sigma
\ .
\label{action}
\end{equation}
where the first term on the R.H.S. is the volume contribution
obtained by integrating on the whole region outside the outer 
horizon, $R$ is the scalar curvature, $\phi$ is the dilaton field, 
$a$ its coupling constant, $F$ is the Maxwell field and $\Sigma$ contains 
all surface terms.
In Ref.~\cite{chlc} the field equations derived from the action in
Eq.~(\ref{action}) were expanded in the charge-to-mass ratio,
$Q/M$, and the perturbative static solution was found, which is
of the form
\begin{eqnarray}
ds^2 = -{\Delta\,\sin^2\theta\over\Psi}\,(dt)^2
+ \Psi\,(d\varphi - \omega\,dt)^2
+\rho^2\,\left[{(dr)^2\over\Delta}+(d\theta)^2\right]
\ .
\label{g_ij}
\end{eqnarray}
The latter can be simplified upon substituting for the (bare)
parameters $M$, $Q$ and $J\equiv\alpha\,M$ the ADM mass, charge and
angular momentum of the hole and also by shifting the radial coordinate,
$r\to r-a^2\,Q^2/6\,M$ (see Ref.~\cite{kndw} for the details).
One finally obtains that the metric in Eq.~(\ref{g_ij}) coincides
(at order $Q^2/M^2$) with the Kerr-Newman solution \cite{chandra}.
This implies that the background dilaton field,
\begin{eqnarray}
\phi=-a\,{r\over\rho^2}\,{Q^2\over M}+{\cal O}(Q^4)
\ ,
\end{eqnarray}
does not affect the causal structure at order
$Q^2/M^2$.
\par
Once analytically continued to the quasi-Euclidean section,
the surface term in Eq.~(\ref{action}) contributes the
surface action of the Kerr-Newman black hole \cite{gh},
\begin{eqnarray}
&&\Sigma={\pi\,M\,(r_+^2+\alpha^2)\over\sqrt{M^2-\alpha^2-Q^2}}
+\Sigma_a
\equiv{\beta_H\over 2}\,M
\ ,
\end{eqnarray}
where $\Sigma_a$ represents (unknown) ${\cal O}(Q^4)$ corrections
from the dilaton, $\beta_H=2\,\pi/\kappa$ is the period of the 
complexified time $T=i\,t$, $\kappa$ being the surface gravity of 
the Kerr-Newman dilaton black hole, and $r_+=M+\sqrt{M^2-\alpha^2-Q^2}$
is the horizon of the Kerr-Newman black hole.
Neither the dilaton nor the electromagnetic field add new surface
contributions because they both fall off fast enough at infinity
and are regular on the horizon, where the measure of
integration $\sqrt{^{(3)}g}\sim\Delta=0$.
\par
The energy-momentum tensor $T^{EM}$ is traceless.
Then, on using Einstein's equations, one can prove that
$R=(\nabla\phi)^2/2$, and the volume contribution to the
action above {\em on shell\/} reduces to
\begin{eqnarray}
S_{EM}&=&-{1\over 16\,\pi}\,
\int d^4x \,\sqrt{g}\,e^{-a\,\phi}\,F^2
\nonumber \\
&=&{\beta_H\over 4}\,\int_{-1}^{+1} d\mu\,
\int_{r_+}^{+\infty} dr\,{1\over\Psi}\,\left[
e^{-a\,\phi}\,\left(\Delta\,A^2_{,r}+\delta\,A^2_{,\mu}\right)
-e^{a\,\phi}\,\left(\Delta\,B^2_{,r}+\delta\,B^2_{,\mu}\right)
\right]
\ .
\end{eqnarray}
The electromagnetic potentials $A$ and $B$ are given by
\cite{chlc,kndw},
\begin{eqnarray}
A&=&Q\,{r\over\rho^2}\,\left[
1-\left({1\over2\, r}+{r\over\rho^2}\right)\,
{a^2\,Q^2\over 3\,M}\right]+{\cal O}(Q^5)
\nonumber \\
B&=&-Q\,\alpha\,{\mu\over\rho^2}\,
\left[1-\left({1\over 2\,M}-{r\over\rho^2}\right)\,
{a^2\,Q^2\over3\,M}\right]+{\cal O}(Q^5)
\ ,
\end{eqnarray}
where $\rho^2=r^2+\alpha^2\,\mu^2$ and the terms proportional to
$Q^2$ inside the brackets are corrections to the Kerr-Newman
potentials.
The integration can be carried out explicitely up to order $Q^4$
to find
\begin{equation}
S_{EM}=-{\beta_H\over 2}\,Q^2\,{r_+\over r_+^2+\alpha^2}
+S_a+{\cal O}(Q^6)
\ ,
\end{equation}
where the first term on the R.H.S. is the contribution from the
Kerr-Newman electromagnetic potentials, and
\begin{equation}
S_a\simeq{\beta_H\over 2}\,{a^2\,Q^4\over 24}\,
{3\,r_+^2+\alpha^2\over M^3\,r_+^2}
\ ,
\label{Sa}
\end{equation}
is a term which vanishes for zero dilaton field.
\par
The total euclidean action for the Kerr-Newman dilaton black hole
is thus given by
\begin{eqnarray}
&&S_{KND}(M,\alpha,Q;a)=
{\beta_H\over 2}\,\left[M-Q^2\,{r_+\over r_+^2+\alpha^2}
+{a^2\,Q^4\over 24}\,{3\,r_+^2+\alpha^2\over M^3\,r_+^2}\right]
+{\cal O}(Q^6)
\ .
\label{knd}
\end{eqnarray}
For $a=0$ one recovers the expression $S_{KN}$ for the 
Kerr-Newman black hole \cite{gh}, which diverges in the {\em extremal\/} 
case $M^2=\alpha^2+Q^2$.
But, as stated above, the quantum degeneracy of states makes use of the 
surface area of the horizon \cite{gh},
\begin{eqnarray}
A_{KN}/4=S_{KN}-\beta_H\,\Omega\,J=\pi\,(r_+^2+\alpha^2)
\ ,
\end{eqnarray}
which is instead well behaved.
When $a\not=0$ and in the limit $|\alpha|\to M$, $S_a$ in Eq.~(\ref{Sa})
still diverges.
Although our knowledge of the metric does not allow us to 
compute $\beta_H$ at order $Q^4$, we can guess that the term 
$-\beta_H\,\Omega\,J$ possibly compensates for such a divergence in the 
same way it was for $S_{KN}$ at order $Q^2$.
This must be so since the surface area of the horizon is finite.
Further, there are hints that the extremal case have zero 
Bekenstein-Hawking entropy.
This is true, {\em e.g.\/}, for the exact dilatonic Reissner-Nordstr\"om 
solution \cite{hh}, for which it is also well known that the extreme 
limit does not commute with the limit of vanishing dilaton parameter 
($a\to 0$).
Therefore it is not clear whether a residual action (at $a=0$) should
be considered in the extremal case.
In this work we shall not make any assumption about the Bekenstein-Hawking 
entropy for the extremal black holes.
\par
According to previous results \cite{hl1}, the statistical mechanics
of a gas of black holes can be consistently formulated only in the
microcanonical ensemble.
The microcanonical density of states for a dilute gas of black holes
described by the Euclidean action in Eq.~(\ref{knd}) of total energy
$E$, total angular momentum $J$ and total charge $Q$ is given by
\begin{equation}
\Omega(E,J,Q;V,a)=\sum\limits_{n=1}^\infty\,\Omega_n(E,J,Q;V,a)
\ ,
\end{equation}
where $V$ is the volume of the system and the density of states for
the configuration with $n$ black holes is
\begin{eqnarray}
\Omega_n(E,J,Q;V,a)&=&
\left[{V\over (2\,\pi)^3}\right]^n\,{1\over n!}\,
\prod\limits_{i=1}^n\,\int_{m_0}^\infty dm_i\,\int_{-m_i^2}^{+m_i^2}
dj_i\,\int_{-\sqrt{m_i^2-\alpha_i^2}}^{+\sqrt{m_i^2-\alpha_i^2}} dq_i\,
\rho_{KND}(m_i,\alpha_i,q_i;a)
\nonumber \\
&&\times\,
\int_{-\infty}^{+\infty} d^3 p_i\,
\delta\left(E-\sum_{i=1}^n E_i\right)\,
\delta\left(Q-\sum_{i=1}^n q_i\right)\,
\delta\left(J-\sum_{i=1}^n j_i\right)\,
\delta^3\left(\sum_{i=1}^n {\bf p}_i\right)
\ ,
\end{eqnarray}
where $\rho_{KND}\sim c\,\exp(A_{KND}/4)$.
We make use of the working assumption that black holes obey the
particle-like dispersion relation $E_i=\sqrt{m_i^2+|{\bf p}_i|^2}$,
where $E_i$ is the energy of the $i^{th}$ black hole with linear
momentum ${\bf p}_i$.
Also, the integrations over the masses $m_i$, the angular momenta $j_i$ 
and the charges $q_i$ are constrained to the domain
$m^2_i\ge \alpha_i^2+q_i^2$, $\forall\,i=1,\ldots,n$.
The mass $m_0\ll M$ is the mass of the least massive black hole
in the gas.
\par
For each $n$, the corresponding density of states can be approximated
by taking the most probable configuration 
${\cal P}_n=\{(m_i',q_i',j_i'), i=1,\ldots,n\}$
which satisfies the constraints expressed by the delta functions in
the integrand above.
First we note that the high linear momentum states contribute
negligibly \cite{hl1}, so that we neglect $|{\bf p}_i|$ with respect
to $m_i$ everywhere and set $M\equiv E$.
Then we argue that ${\cal P}_n$ does not satisfy the
equipartition principle, that is ${\cal P}_n\not=
{\cal E}_n\equiv\{(m_i=M/n,q_i=Q/n,j_i=J/n), i=1,\ldots,n\}$.
\par
\begin{figure}
\centerline{\epsfysize=300pt\epsfbox{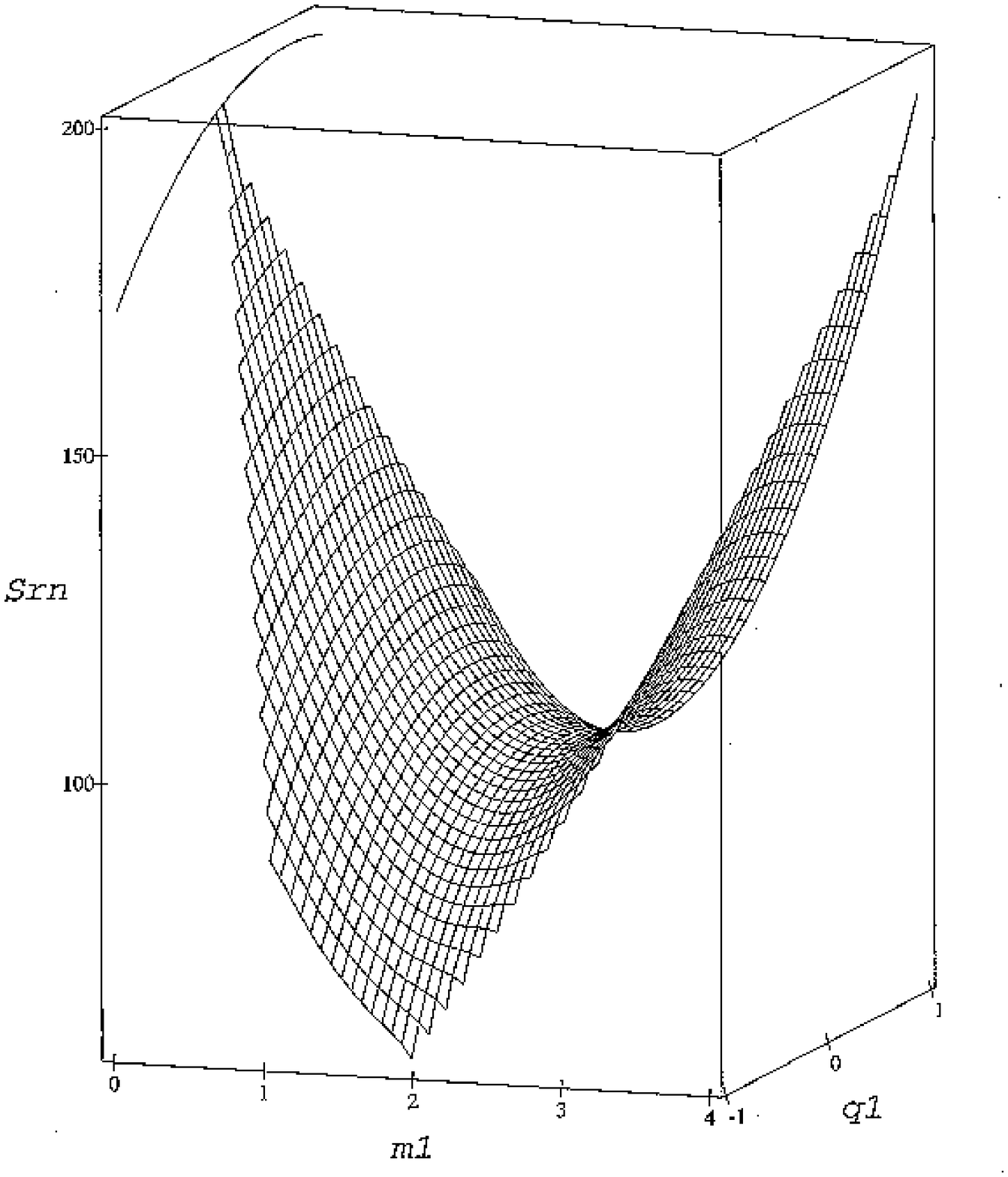}}
\caption{The total Bekenstein-Hawking entropy $S_{rn}$ for a system 
of two Reissner-Nordstr\"om black holes with total mass $M=4$ and 
total charge $Q=1$ as a function of $m_1$ and $q_1$.}
\label{a}
\end{figure}
\begin{figure}
\centerline{\epsfysize=300pt\epsfbox{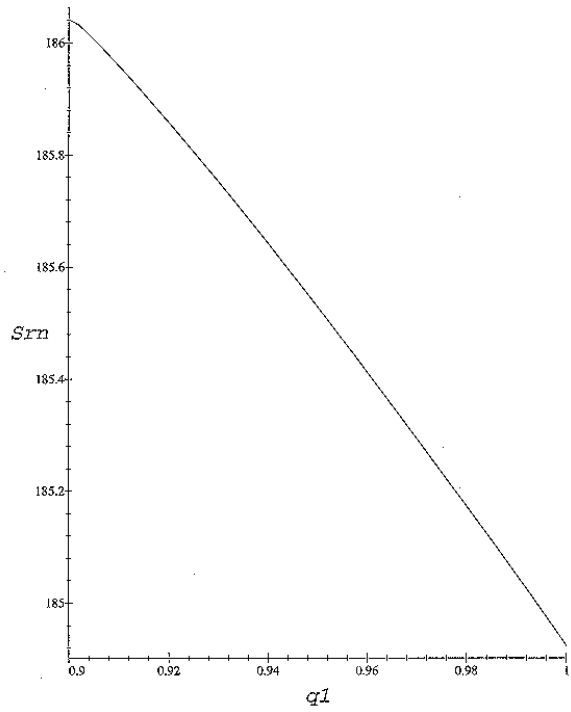}}
\caption{When the lower limit for the mass of each black hole is 
$m_0=0.5$, the action $S_{rn}$ in Fig.~\ref{a} has a maximum for 
$m_1=4-m_0$ and $q_1=1-m_0$, meaning that the second black hole 
is extremal ($m_2=q_2=m_0$).}
\label{b}
\end{figure}

It is possible to prove the statement above for a gas of $n$ 
Reissner-Nordstr\"om black holes ($a=j_i=0$, $i=1,\ldots,n$),
whose total Bekenstein-Hawking entropy is given by the sum 
\begin{equation}
S_{tot}(M,Q)=\sum\limits_{i=1}^n\,S_{RN}(m_i,q_i)
\ ,
\end{equation}
together with the constraints $\sum_{i=1}^n\,m_i=M$, $\sum_{i=1}^n\,q_i=Q$,
$m_i\ge |q_i|$, $i=1,\ldots,n$, and 
$S_{RN}=\pi\,(M+\sqrt{M^2-Q^2})^2$.
By using the Lagrange multiplier technique, one finds that the extrema
of $S_{tot}$ are given by solutions of the following equations
\begin{equation}
\left\{\begin{array}{l}
r^2_{i+}=\lambda_m\,\sqrt{m_i^2-q_i^2} \\
r_{i+}\,q_i=-\lambda_Q\,\sqrt{m_i^2-q_i^2}
\end{array}\right.
\ \ \ \ \ i=1,\ldots,n
\ ,
\label{lag}
\end{equation}
where $r_{i+}=m_i+\sqrt{m_i^2-q_i^2}$ and $\lambda_m$, $\lambda_Q$ are 
Lagrange multipliers corresponding respectively to the mass and charge 
constraint.
These are algebraic equations of order four, thus one expects a certain 
number of different solutions to be available for each black hole in 
the configuration.
However, since the black holes in the gas are supposed to interact 
negligibly with each others, the equations in Eq.~(\ref{lag}) decouple 
in the index $i$.
Further, they are exactly the same for all $i=1,\ldots,n$ and one 
solution certainly exists which corresponds to equipartition of mass and 
charge.
But this extremum is not a maximum, since the corresponding
total Bekenstein-Hawking entropy satisfies
\begin{equation}
S_{tot}({\cal E}_n)
=\sum\limits_{i=1}^n\,S_{RN}(M/n,Q/n)
={1\over n}\,S_{RN}(M,Q)
\ ,
\end{equation}
where $S_{RN}(M,Q)$ is also the total Bekenstein-Hawking entropy for a 
configuration in which one black hole carries all the mass and charge
(assuming $m_0=0$ and $M>|Q|$).
Indeed, for $n=2$ we are able to show graphically that $S_{RN}(M,Q)$
is actually the absolute maximum $S_{tot}({\cal P}_n)$, while ${\cal E}_n$ 
is only a saddle point (see Fig.~\ref{a}).
Moreover, if $m_0\not=0$, the most favored configuration is the one in which
the light black hole is extremal, $q_2=m_2=m_0$ (see Fig.~\ref{b}).
We are presently carrying on a numerical study which seems to support
the conjecture that this is the case $\forall\,n>1$ and that one has
${\cal P}_n=\{(M-(n-1)\,m_0,Q-(n-1)\,m_0),
(m_i'=m_0,q_i'=m_0), i=2,\ldots,n\}$,
where the choice of $i=1$ being the most massive black hole is
arbitrary.
\par
For $a=0$ but $j_i\not\equiv 0$, the proof that ${\cal E}_n$ does not
extremize the entropy follows along the same lines.
One just needs to notice that the equations defining the possible
extrema of the total Bekenstein-Hawking entropy of a gas of 
Kerr-Newman black holes,
\begin{equation}
{1\over 4}\,A_{tot}(M,\alpha,Q;a)={1\over 4}\,\sum\limits_{i=1}^n\,
A_{KN}(m_i,\alpha_i,q_i)
\ ,
\end{equation}
are still of the kind in Eq.~(\ref{lag}) and do not contain cross
terms in the index $i$.
However, $A_{KN}(M/n,J/n,Q/n;a)$ is not well defined for
$n>J/M^2$, since one would have $m_i^2=(M/n)^2<(J/M)^2=\alpha_i^2$.
Thus, in general, ${\cal E}_n$ is not even an acceptable configuration.
Further, a numerical analysis for $n=2$ shows that
${\cal P}_n\simeq\{(M-m_0,J,Q-m_0),(m_0,0,m_0)\}$ for $M>|J/M|$, $M>|Q|$.
By setting $q_1=q_2=0$ one obtains plots of $A_{tot}$ as function 
of $m_1$ and $j_1$ which look qualitatively the same as the ones 
in Figs.~\ref{a},~\ref{b}.
\par
If $A_{KND}$ is regular in the limit $|\alpha|\to M$, then in the general
case one is led to consider a configuration with one massive black 
hole which carries most of the charge and angular momentum and is 
surrounded by $n-1$ lighter, extremal black holes.
Then the density of states can be approximated by
\begin{eqnarray}
\Omega_n(M,J,Q;V,a)&\simeq&\left[{c\,V\over(2\,\pi)^3}\right]^n\,
{1\over n!}\,
e^{{n-1\over 4}\,A_{KND}(m_0,\gamma\,m_0,\sqrt{1-\gamma^2}\,m_0;a)}
\nonumber \\
&&\times\, e^{{1\over 4}\,A_{KND}
\left(M-(n-1)\,m_0,J-(n-1)\gamma\,m_0^2,Q-(n-1)\,\sqrt{1-\gamma^2}\,m_0;
a\right)}
\ ,
\end{eqnarray}
where $0\le\gamma\le 1$.
Preliminary numerical calculations seem to suggest that $\gamma\ll 1$.
We intend to perform a complete numerical treatment for several (large)
values of $n$ in a future publication.
\par
Now we can determine the most probable number $N$ of black holes in the
gas, for which $d\Omega_n/dn|_{n=N}=0$, and further approximate
\begin{equation}
\Omega(M,J,Q;V,a)\simeq \Omega_N(M,J,Q;V,a)
\ .
\label{omega}
\end{equation}
\par
We can now check whether the gas of black holes we have been describing
satisfies the bootstrap condition \cite{hag},
\begin{equation}
\lim\limits_{M\to\infty}\,{\Omega(M,J,Q;V,a)\over
\rho_{KND}(M,J,Q;a)}
=1
\ .
\label{boots}
\end{equation}
Indeed, it does, provided $m_0=0$ and
\begin{equation}
e^{N\,\Psi(N+1)}/N!
=c
\label{c}
\ .
\end{equation}
As in the case of a gas of Schwarzschild black holes \cite{hl1},
this equation relates the constant $c$ to the volume $V$.
The number $N$ is then given by $\Psi(N+1)\simeq\ln[c\,V/(2\,\pi)^3]$,
where $\Psi$ is the psi function.
Correspondingly, one obtains the inverse microcanonical temperature
$\beta=d\ln\Omega/dE\simeq d\ln\Omega_N/dM$,
\begin{eqnarray}
\beta&=&\beta_H
\left(M-(N-1)\,m_0,J-(N-1)\gamma\,m_0^2,Q-(N-1)\,\sqrt{1-\gamma^2}\,m_0;
a\right)
\ ,
\end{eqnarray}
which gives exactly the Hawking temperature $1/\beta_H$ when the
bootstrap condition ($m_0=0$) is satisfied. 
\par
Our results show that the equilibrium state for a gas of Kerr-Newman 
dilaton black holes is very far from thermal equilibrium.
Not only does one black hole acquire nearly all of the mass as in the 
Schwarzschild case, but it also acquires most of the charge and of 
the angular momentum of the whole gas, the other black holes
in the gas being much lighter and extremal.
Further, when the mass of the lighter black holes vanishes,
the bootstrap condition is satisfied at high energy.  
Thus the interpretation of the inverse of the tunneling probability 
as obtained from the WKB approximation as the quantum mechanical 
degeneracy of states rather than as the statistical mechanical density 
of states holds for a gas of black holes each of whose members may
be given arbitrary mass, charge and angular momentum in some initial 
state.
Of course, the final equilibrium configuration of the gas is given
by the inhomogeneous distribution described above.


\end{document}